\documentclass[aps,preprint]{revtex4}
\usepackage{amsfonts}
\usepackage{amsmath}
\usepackage{makeidx}
\usepackage{amsmath}
\usepackage{graphicx}
\usepackage{epstopdf}
\usepackage{amssymb,epsf}

\begin{document}

\title{Magnetic Brane of Cubic Quasi-Topological Gravity in the
Presence of Maxwell and Born-Infeld Electromagnetic Field}
\author{ M. Ghanaatian $^{1}$ \footnote{%
email address: $m_{-}ghanaatian@pnu.ac.ir$}, A. Bazrafshan $^{2}$,
S. Taghipoor $^{1}$ and R. Tawoosi $^{1}$}

\affiliation{$^1$ Department of Physics, Payame Noor University,
PO BOX 19395-3697 Tehran, Iran } \affiliation{$^2$ Department of
Physics, Jahrom University, 74137-66171 Jahrom, Iran}

\begin{abstract}
The main purpose of the present paper is analyzing magnetic brane
solutions of cubic quasi-topological gravity in the presence of a
linear electromagnetic Maxwell field and a nonlinear
electromagnetic Born-Infeld field. We show that the mentioned
magnetic solutions have no curvature singularity and also no
horizons, but we observe that there is a conic geometry with a
related deficit angle. We obtain the metric function and deficit
angle and consider their behavior. We show that the
attributes of our solution are dependent on cubic
quasi-topological coefficient and the Gauss-Bonnet parameter.
\end{abstract}

\maketitle

%\pacs{04.50.+h, 04.20.Jb, 04.70.Bw, 04.70.Dy}

\section{Introduction}

This is great interest in finding the horizonless solutions in
various theories of gravity, because these kinds of solutions may
be interpreted as cosmic strings/branes. The cosmic strings/branes
are topological defects which were formed during phase transitions
in the early universe. The cosmic strings/branes are known as
super-conducting strings. Witten \cite{Witten} showed that these
strings behave like super-conductors and interact with the
astronomical magnetic field. The cosmic strings/branes are able to
generate strong astrophysical magnetic fields
\cite{Vachaspati1,Brandenberger}. Therefore, the string model of
structuring can help to disclose the origin of cosmic magnetic
fields \cite{Vachaspati2}. Besides cosmological roles
\cite{Hansen,Bezerra1,Bezerra2,Bezerra3}, they are fascinating
objects, because they have no curvature singularity and no
horizon, but they have a conic singularity with a deficit angle.
These solutions for space-times which have flat hyper-surfaces of
$t=constant$ and $r=constant$, have no curvature  singularity and
no horizon, but they have a conic singularity.

Despite the marvelous success of Einstein's general relativity,
some observations, such as the accelerated expansion of the
universe, galactic rotation curves and so on, are strongly
influenced by the re-emergence of modified gravitation theories.
Recently, extensive efforts have been made to generalize the
Hilbert-Einstein Lagrangian based on the presence of gravitational
fields near intrinsic singularities and the creation of a
first-order approximation in quantum theory of gravitation. One of
the generalized theories of gravity is gravity with higher-order
terms of curvature tensor. The Hilbert-Einstein Lagrangian is
linear in relation to the curvature tensor, but there is such a
feeling that there was no initial reason for such a particular
choice. After the presentation of general relativity, attempts
were made to quantize the gravitational field. In the meantime,
super-gravity theories were put forward, which in these theories,
the divergences were not completely eliminated. The search for
quantum gravity has made the theory of strings in which the
particle model is not considered as a point, but is similar to a
one-dimensional object called string. Some gravitons were proposed
as candidates for gravitational quantum theory. The combination of
super-symmetry and gravity in these theories results in
super-string theory \cite{Leigh}, which is consistent in ten
dimensions. One of super-string theories, which are so much
considered, is the Theory of everything (TOE). The other reason
for using high-order gravities is that, if the curvature of space
is not noticeable, Einstein's gravity as the basis of quantum
gravity is not a normalized theory, and therefore, terms
containing higher curvatures should be added to the
Einstein-Hilbert Lagrangian. What is presented in the cosmological
branes theory tends to increase gravity in higher dimensions. In
these theories, space-time dimensions are considered to be higher
than four and gravity is supposed to be higher in dimensions, and
other materials and non-gravitational interactions can be immersed
in a 4-dimensional approach \cite{Aoyanagi}. These theories should
be reduced to Einstein's gravity in the low energy field. In the
four-dimensional space, the only action that leads to the
second-order equations of the metric is the Einstein-Hilbert
action. But in higher dimensions, the actions that
lead to the second-order equations can be introduced. One of these theories is
Lovelock's gravity \cite{Lovelock1,Lovelock2}. The Lovelock theory
is the most general theory which was proposed to have field
equation with second order derivatives of metric. In recent years,
the gravity of Lovelock has been of great interest. Many writers
have studied the gravity of Lovelock up to second order, which is
known as the Gauss-Bonnet gravity
\cite{Fradkin,Choquet,Brigante,Ge,Cai,Hu}. It was the most general
Lagrangian in the 5 and 6 dimensions. The third order of Lovelock
gravity has attracted many physicists. The reason for this is that
the holography of the third-order Lovelock theory has four
coupling constants, which makes it possible to make a larger group
of homogeneous fields in holographic terms. The Lagrangian of
third-order Lovelock theory is the general Lagrangian in seven and
eight dimensions whose field equations have maximum second order
derivatives of metric. One may note that the cubic term of
Lovelock gravity has no contributions in five dimensions, while
the cubic term of quasi topological gravity introduced in
\cite{Myers} has contribution in five dimensions. So, in this
paper, we consider the cubic term of quasi topological gravity.

In 1934, Born and Infeld \cite{BI} presented a nonlinear
electrodynamics theory to get rid of the unlimited self-energies
of point-like charged particles such as electrons. In the limit of
weak fields, the Born-Infield Lagrangian is reduced to the Maxwell
Lagrangian and a series of small corrections. In recent years,
with the development of mathematics as well as the development of
super-string theory, the Born-Infield theory was very much taken
into consideration. For example, the dynamics of the D-branes and
some of soliton solutions of super-gravity are based on the
Born-Infeld theory. Born-Infeld theory is now associated with the
theory of non-Abelian super-symmetry, noncommutative geometry,
Cayley-Dickson algebra, and other physics-related disciplines.
Recently, quantum cosmology has been studied along with the Scalar
born-Infield fields. Also, the classical solutions of the wormhole
and wormhole wave function have been obtained along with the
Born-Infeld scalar fields, and their interesting characteristics
have been investigated.

The extension of magnetic solutions to the higher curvature
gravity with the linear and nonlinear electromagnetic field has
also been done
\cite{Dehghani1,Dehghani2,Hendi1,Hendi2,Hendi3,Hendi4,Hendi5,Hendi6,Hendi7}.
In this paper, we want to obtain (n + 1)-dimensional magnetic
solutions of cubic quasi-topological gravity in the presence of
the linear and nonlinear electromagnetic field and consider the
effects of the electromagnetic Maxwell and Born-Infeld field on
the details of the space-time such as the deficit angle of the
space-time. The outline of our paper is as follows: We bring a
brief review of cubic quasi-topological action in the presence of
the electromagnetic field in Sec. \ref{Action}. Sec.
\ref{Solutions} will start with presenting the metric for the
static horizonless solutions. Using this metric, we calculate the
magnetic solution of cubic Quasi-Topological-Maxwell gravity in
Subsec. \ref{Solutions1} and analyze the magnetic brane of cubic
Quasi-Topological-Born-Infeld gravity in Subsec. \ref{Solutions2}.
In these two subsections, the behavior of the metric function and
the deficit angle are taken into account and the effects of the
quasi-topological coefficient and the Gauss-Bonnet parameter are
obtained. Finally, we finish our research with some concluding
remarks in Sec. \ref{conclusion}.

\section{\ Quasi-Topological Action \label{Action}}

Here, we will present the action of quasi-topological gravity up
to third order in $(n+1)$ dimensions in the presence of an
electromagnetic field as follows
\cite{Dehghani,Bazrafshan,Ghanaatian1,Ghanaatian2,Ghanaatian3}:
\begin{equation}
I_{G}=\frac{1}{16\pi }\int d^{n+1}x\sqrt{-g}[-2\Lambda
+\mathcal{L}_{1}+\mu_{2}\mathcal{L}_{2}+\mu_{3}\mathcal{X}_{3}+L(F)].
\label{Act01}
\end{equation}
where $\Lambda =-n(n-1)/2l^{2}$ is the cosmological constant, $\mathcal{L}_{1}={R}$ is just the Einstein-Hilbert Lagrangian, $%
\mathcal{L}_{2}=R_{abcd}{R}^{abcd}-4{R}_{ab}{R}^{ab}+{R}^{2}$ is
the second order Lovelock (Gauss-Bonnet) Lagrangian,
$\mathcal{X}_{3}$\ is the curvature-cubed Lagrangian of the
quasi-topological gravity
\cite{Myers,Dehghani,Bazrafshan,Ghanaatian1,Ghanaatian2,Ghanaatian3}:
\begin{eqnarray}
\mathcal{X}_{3} &=&R_{ab}^{cd}R_{cd}^{\,\,e\,\,\,f}R_{e\,\,f}^{\,\,a\,\,\,b}+%
\frac{1}{(2n-1)(n-3)}\left(
\frac{3(3n-5)}{8}R_{abcd}R^{abcd}R\right.
\notag \\
&&-3(n-1)R_{abcd}R^{abc}{}_{e}R^{de}+3(n+1)R_{abcd}R^{ac}R^{bd}  \notag \\
&&\left. +\,6(n-1)R_{a}{}^{b}R_{b}{}^{c}R_{c}{}^{a}-\frac{3(3n-1)}{2}%
R_{a}^{\,\,b}R_{b}^{\,\,a}R+\frac{3(n+1)}{8}R^{3}\right) .
\label{X3}
\end{eqnarray}
and $ L(F)$ is an arbitrary Lagrangian of the electromagnetic
field, where we use the maxwell lagrangian as a linear
electromagnetic field in the subsection (\ref{Solutions1}) and the
Born-Infeld lagrangian as a nonlinear electromagnetic field in the
subsection (\ref{Solutions2}). Note that $\mathcal{X}_{3}$\ is
only effective in dimensions greater than four and they become
trivial in six dimensions
\cite{Myers,Dehghani,Bazrafshan}.

\section{STATIC MAGNETIC BRANES \label{Solutions}}

In this section, we want to obtain the solutions of the cubic
quasi-topological gravity in the presence of a linear and
nonlinear electromagnetic field. We will work with the following
metric \cite{Dias1,Dias2,Dias3}:
\begin{equation}
ds^{2}=-\rho ^{2}/l^{2}dt^{2}+\frac{d\rho ^{2}}{f(\rho
)}+l^{2}{g(\rho )}d\phi ^{2}+\frac{\rho
^{2}}{l^{2}}\sum\limits_{i=1}^{n-2} d\theta _{i}^{2} \label{met01}
\end{equation}
where $l$ is a scale factor related to the cosmological constant
and $\sum\limits_{i=1}^{n-2} d\theta _{i}^{2}$ is the Euclidean
metric. Using this metric, we want to obtain the magnetic
solutions with no horizon. Therefore, instead of using
Schwarzshild metric $[(g_{\rho\rho})^{-1}\propto (g_{tt})$ and $(
g_{\phi\phi}) \propto \rho^{2}]$, we use the metric like
$[(g_{\rho\rho})^{-1}\propto (g_{\phi\phi})$  and $( g_{tt})
\propto -\rho^{2}]$. In this metric, $f(\rho)$ and $g(\rho)$ are
arbitrary functions of $\rho$ (the radial coordinate) and we
should find the values of them. Here, $\phi$ is the angular
coordinate and it ranges in $0\leq\phi<2\pi$ and it is
dimensionless.

\subsection{The Magnetic Solutions of Quasi-Topological-Maxwell Gravity  \label{Solutions1}}

By using the metric (\ref{met01}), we can obtain the horizonless
solutions that are of our interest. First, we want to obtain the
solution of quasi-topological gravity in the presence of the
linear maxwell electromagnetic field. The lagrangian of maxwell
electromagnetic field is
\begin{equation}
L(F)=-F^{2}
\end{equation}
where $F^{2}=F_{\mu \nu }F^{\mu \nu }$ is the maxwell invariant,
$F_{\mu \nu }=\partial _{\mu }A_{\nu }-\partial _{\nu }A_{\mu }$
is the electromagnetic field tensor and $A_{\mu } $ is the vector
potential. Using the metric (\ref{met01}) and
\begin{equation}
A_{\phi }=h(\rho ).
\end{equation}%
for the vector potential, we can obtain the below action per unit
volume by integrating by parts as
\begin{equation}
I_{G}=\frac{{(n-1)}}{16\pi l^{2}}\int dtd\rho \{ N(\rho) \left[
\rho ^{n}(1+\xi +\hat{\mu}_{2}\xi ^{2}+\hat{\mu}_{3}\xi
^{3})\right]'+\frac{2l^2\rho ^{n-1}{h'}^{2}}{N(\rho)(n-1)} \}.
\label{Act02}
\end{equation}
where $\xi =-l^{2}\rho^{-2}f(\rho)$, $g(\rho)=N(\rho)^{2}f(\rho)$ and the dimensionless parameters $%
\hat{\mu}_{2}$, $\hat{\mu}_{3}$ are defined as:
\begin{equation*}
\hat{\mu}_{2}\equiv \frac{(n-2)(n-3)}{l^{2}}\mu_2,\text{ \ \ \ }\hat{\mu}%
_{3}\equiv \frac{(n-2)(n-5)(3n^{2}-9n+4)}{8(2n-1)l^{4}}\mu_3,
\end{equation*}%

\begin{figure}[htbp]
    \includegraphics[scale=0.8]{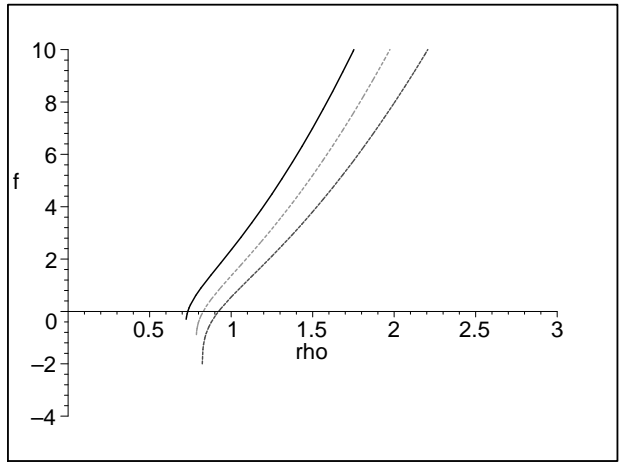}
    \caption{The overlay plot of $f(\rho)$ versus $\rho$ for $\hat{\mu}_{3}=0.008$
    (solid), $\hat{\mu}_{3}=0.009$ (dotted) and $\hat{\mu}_{3}=0.01$ (dashed). Here, $l =1
    $, $\hat{\mu}_{2}= 0.1$, $q=10$ and $m=0.1$.}
    \label{max3f01}
\end{figure}

\begin{figure}[htbp]
    \includegraphics[scale=0.8]{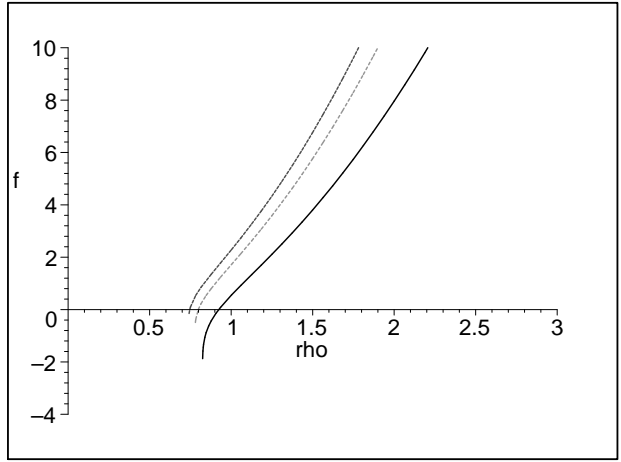}
    \caption{The overlay plot of $f(\rho)$ versus $\rho$ for $\hat{\mu}_{2}= 0.1$
    (solid), $\hat{\mu}_{2} = 0.11$ (dotted) and $\hat{\mu}_{2} = 0.115$ (dashed). Here, $l =1%
    $, $\hat{\mu}_{3}=0.01$, $q=10$ and $m=0.01$.}
    \label{max3f02}
\end{figure}

Varying the action (\ref{Act02}) with respect to $\xi (\rho)$
yields
\begin{equation}
\left( 1+2\hat{\mu}_{2}\xi +3\hat{\mu}_{3}\xi
^{2}\right) \frac{dN(\rho )}{d\rho }=0,
\label{eom1}
\end{equation}%
which shows that $N(\rho)$ should be a constant. Variation with
respect to $h(\rho)$ and substituting $N(\rho)=1$ gives
\begin{equation}
(n-1) h'+\rho h''=0,
\label{eom2}
\end{equation}%
So, we can calculate the vector potential as
\begin{equation}
h(\rho)=-\frac{q}{\rho^{n-2}}, \label{Amu0}
\end{equation}
where $q$  is related to the charge parameter which is an
integration constant. Variation with respect to $N(\rho)$ and
substituting $N(\rho)=1$ gives
\begin{equation}
\hat{\mu}_{3}\xi ^{3}+\hat{\mu}_{2}\xi^{2}+\xi +\kappa =0,
\label{Fr01}
\end{equation}%
where
\begin{equation}
\kappa
=1-\frac{m}{\rho^{n}}+\frac{2(n-2)l^{2}q^{2}}{\rho^{2(n-1)}(n-1)}
\end{equation}%
and $m$ is an integration constant which is like the mass of the
space-time. We should solve the cubic equation (\ref{Fr01}) to
find the value of $f(\rho)$. By substituting
$\xi=J-\frac{\hat{\mu}_{2}}{3\hat{\mu}_{3}}$, the equation
(\ref{Fr01}) becomes:
\begin{equation}
J^3+\alpha_{1}J+\alpha_{2}=0  \label{Fr4}
\end{equation}
where
\begin{equation}
\alpha_{1}=\frac{-\hat{\mu}_{2}^{2}+\hat{\mu}_{3}}{\hat{\mu}_{3}^{2}}
\end{equation}
\begin{equation}
\alpha_{2}=\frac{-\kappa\hat{\mu}_{2}^{2}-\hat{\mu}_{2}\hat{\mu}_{3}+4\hat{\mu}_{2}^{3}}{27\hat{\mu}_{3}^{3}}
\end{equation}
When $\Delta=\alpha_{1}^{2}-\alpha_{2}^{3}>0$ then the Eq.
(\ref{Fr4}) has a real root:
\begin{equation}
\omega_{1}=(\alpha_{1}+\sqrt{\alpha_{1}^{2}-\alpha_{2}^{3}})^{\frac{1}{3}}
\end{equation}
\begin{equation}
\omega_{2}=(\alpha_{1}-\sqrt{\alpha_{1}^{2}-\alpha_{2}^{3}})^{\frac{1}{3}}
\end{equation}
So we have:
\begin{equation}
\xi_{1}=\omega_{1}+\omega_{2}-\frac{\hat{\mu}_{2}}{3\hat{\mu}_{3}}
\end{equation}
\begin{equation}
f(\rho)=-l^{2}\rho^{-2}\xi_{1} \label{Fr}
\end{equation}
Figures (\ref{max3f01}) and (\ref{max3f02}) indicate that
$f(\rho)$ has a root that we call it as $r_{+}$ and it seems that
there is a curvature singularity at $\rho=r_{+}$. In Fig.
(\ref{max3f01}), when we increase the coefficient value of
quasi-topological gravity, $\hat{\mu}_{3}$, the value of $r_{+}$
increases. And it becomes clear that the metric function,
$f(\rho)$, is positive for the large value of $\rho\gg r_{+}$.
Fig. (\ref{max3f02}) shows  that increasing value of the GB
parameter, $\hat{\mu}_{2}$, leads to decreasing $r_{+}$.

For studying the general properties of the solution given in Eq.
(\ref{Fr}), at first we look for the main curvature singularities.
We can see easily that the Kretschmann scalar $R_{abcd}R^{abcd}$
diverges at $\rho= 0$, and it may seem there is a curvature
singularity at $\rho= 0$. In the following, we look for the
horizon for the main singularity. If we have horizon, the function
$f(\rho )$ becomes zero at the radius of horizon. Suppose that
$r_+$  is the largest real root of $f(\rho ) = 0$. So, the
function $f(\rho )$  is negative for $\rho < r_+$ and positive for
$\rho > r_+$. But, we see the space-time will never achieve $\rho=
0$ and this analysis is not true. We denote the metric signature,
in the range  $0 < \rho < r_+$ , may change from $(- + + + +...+)$
to $(- - - + +...+)$. By accounting this apparent change of
signature of the metric, we conclude that we can not extend the
space-time to $\rho < r_+$. So, we do the following suitable
transformation to get ride of this incorrect extension by
introducing a new radial coordinate r:
\begin{equation}
r=\sqrt{{\rho }^{2}-{r_{+}}^{2}}\Rightarrow d{\rho }^{2}=\frac{r^{2}}{%
r^{2}+r_{+}^{2}}dr^{2} \label{trans01}
\end{equation}
Then by this transformation the value of  the metric \ref{met01}
become

\begin{equation}
ds^{2}=-\frac{(r^{2}+r_{+}^{2})}{l^{2}}dt^{2}+\frac{{r^{2}dr^{2}}}{(r^{2}+r_{+}^{2})f(r)}%
+l^{2}g(r)d\phi^{2}+\frac{(r^{2}+r_{+}^{2})}{l^{2}}\sum\limits_{i=1}^{n-2}
d\theta _{i}^{2}. \label{met02}
\end{equation}%
In this metric, the range of $\phi$ and $r$ are $0\leq\phi<2\pi$
and $0\leq r <\infty$, respectively. So the electrodynamic field
and the metric functions are real for $r\geq0$. In addition, the
function $f(r)$ is positive in the whole space-time and is zero at
$r=0$. The kretschmann scalar does not diverge in the range $
0\leq r <\infty$, but we can show that there is a conical
singularity at $r=0$. Therefore, we use the Taylor expansion in
the vicinity of $r=0$ as following:
\begin{equation}
f(r)=f(r)|_{r=0}+(\frac{df(r)}{dr}|_{r=0})r+\frac{1}{2}(\frac{d^{2}f(r)}{d^{2}r}|_{r=0})r^{2},
+O(r^{3})+...,
\end{equation}%
where
\begin{equation}
f(r)|_{r=0}=\frac{df(r)}{dr}|_{r=0}=0,
\end{equation}%
And by this transformation (\ref{trans01}), the functions $h(r)$
and $\kappa$ become
\begin{equation}
h(r)=-\frac{q}{{(r^{2}+r_{+}^{2})}^{{(n-2)}/2}}, \label{Amu1}
\end{equation}
\begin{equation}
\kappa
=1-\frac{m}{{(r^{2}+r_{+}^{2})}^{n/2}}+\frac{2(n-2)l^{2}q^{2}}{{(n-1)(r^{2}+r_{+}^{2})}^{(n-1)}}
\end{equation}%

We can investigate the conic geometry by using the circumference/radius ratio:
\begin{equation}
{\lim}_{\left(r\rightarrow0
\right)}{\left(\frac{1}{r}{\sqrt{\frac{g_{\phi\phi}}{g_{rr}}}}
\right)}\neq1 \label{limit}
\end{equation}
when the radius $r$ tends to zero, the limit of the ratio
\textquotedblleft circumference/radius\textquotedblright\ is not
$2\pi$, so we can conclude near $r = 0$, the metric describes a
spacetime that is locally flat and has a conical singularity at $r
= 0$ with a deficit angle $\delta\phi= 8\pi\tau$. We can remove
the conical singularity if we identify the coordinate $\phi$ with
the period
\begin{equation}
period_\phi=2\pi{\lim}_{\left(r\rightarrow0
\right)}{\left(\frac{1}{r}{\sqrt{\frac{g_{\phi\phi}}{g_{rr}}}}
\right)}^{-1}=2\pi(1-4\tau) \label{period}
\end{equation}
That $\tau$ is
\begin{equation}
\tau=\frac{1}{4}\left(1-\frac{2}{lr_+f''_{0}}\right) \label{tau}
\end{equation}
In above equation $f''_{0}$ is the value of the second derivative of
$f(r)$ at $r=0$.
From the above analysis, we can conclude that near the origin, $r
= 0$, the metric (\ref{met02}) may be written as
\begin{equation}
ds^2=\frac{r_+^{2}}{l^{2}}\left(-dt^{2}+\sum\limits_{i=1}^{n-2}
d\theta _{i}^{2}\right)+\frac{dr^{2}}{r_+^{2}f^{''}}+l^{2}r_+^{2}f^{''}d\phi^{2}
\end{equation}

\begin{figure}[htbp]
    \includegraphics[scale=0.8]{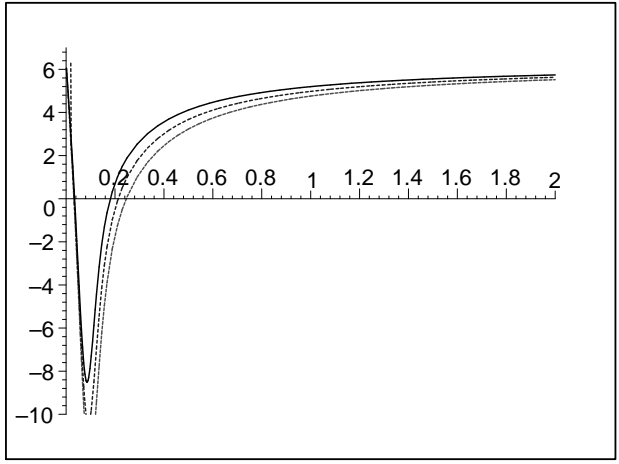}
    \caption{The overlay plot of $\delta$ versus $r_+$ for $\hat{\mu}_{3}=1$
    (solid), $\hat{\mu}_{3}=1.2$ (dotted) and $\hat{\mu}_{3}=1.4$ (dashed). Here, $l =1%
    $, $\hat{\mu}_{2}= 10$ and $q=0.2$.}
    \label{max3dc01}
\end{figure}

\begin{figure}[htbp]
    \includegraphics[scale=0.8]{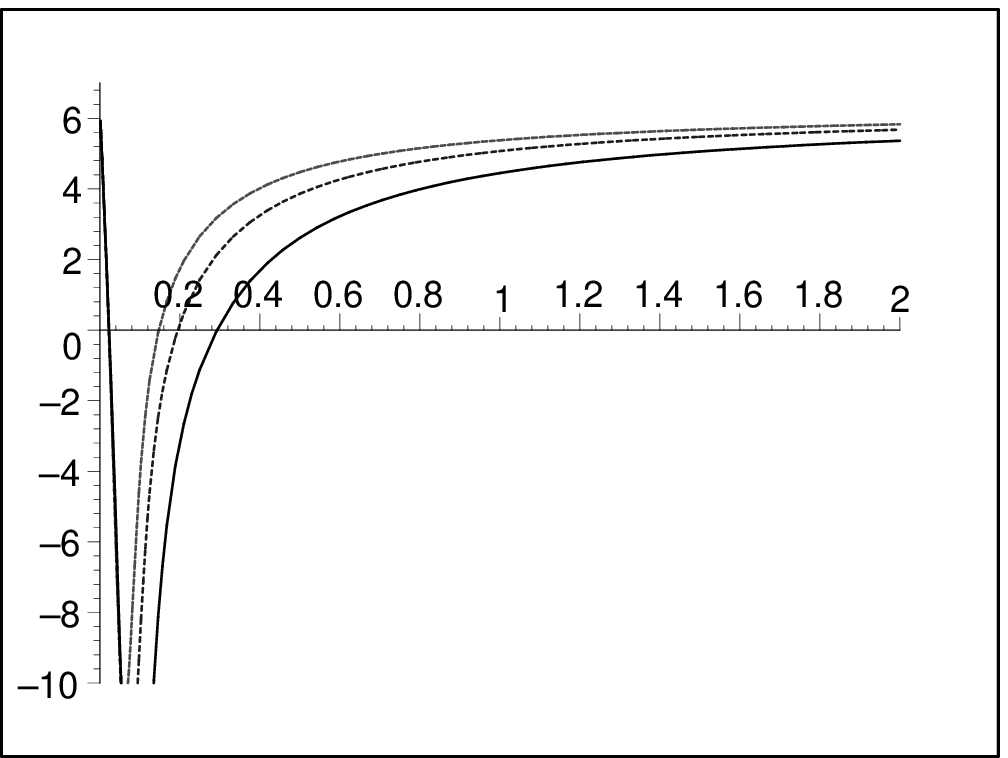}
    \caption{The overlay plot of $\delta$ versus $r_+$ for $\hat{\mu}_{2}=6$
    (solid), $\hat{\mu}_{2}=9$ (dotted) and $\hat{\mu}_{2}=12$ (dashed). Here, $l =1%
    $, $\hat{\mu}_{3}= 1$ and $q=0.1$.}
    \label{max3dc02}
\end{figure}

Then, we check the effects of different parameters of
quasi-topological action on the deficit angle of the space-time.
For this purpose, we plot $\delta$ versus the parameter $r_+$.
This is shown in Figures (\ref{max3dc01}) and (\ref{max3dc02})
which find that the deficit angle $\delta$ is an increasing
function of $r_+$. In Fig. (\ref{max3dc01}), one can find $r_{+}$
increases as the $\hat{\mu}_{3}$ parameter of the
quasi-topological action increases, whereas in Fig.
(\ref{max3dc02}), for increasing the $\hat{\mu}_{2}$ parameter of
Gauss-Bonnet action, $r_{+}$ decreases.

\subsection{The Magnetic Solution of Quasi-Topological-Born-Infeld Gravity \label{Solutions2}}
Here we use the metric (\ref{met01}) and the lagrangian of the
electromagnetic Born-Infeld field as
\begin{equation}
L(F)=4\beta ^{2}\left( 1-\sqrt{1+\frac{F^{2}}{2\beta ^{2}}%
}\right) \label{LBI}.
\end{equation}
where $F^{2}=F_{\mu \nu }F^{\mu \nu }$, $F_{\mu \nu }=\partial _{\mu
}A_{\nu }-\partial _{\nu }A_{\mu }$ is the electromagnetic field
tensor and $A_{\mu } $ is the vector potential. One may note that
in the limit $\beta \longrightarrow\infty$ reduces to the standard
Maxwell form $L(F)=-F^{2}$. By substituting the metric
(\ref{met01}) and the value of $L(F)$ as the lagrangian of the
electromagnetic Born-Infeld field in the action (\ref{Act01}), we
have
\begin{equation}
I_{G}=\frac{{(n-1)}}{16\pi l^{2}}\int dtd\rho \lbrack N(\rho ) {{{\left[
\rho ^{n}(1+\xi +\hat{\mu}_{2}\xi ^{2}+\hat{\mu}_{3}\xi
^{3})\right] ^{\prime }+\frac{4l^{2}\beta^{2}\rho ^{(n-1)}(1-\sqrt{1-\frac{h^{\prime 2}}{\beta^{2}})}}{%
N(\rho )(n-1)}}}}].  \label{Act03}
\end{equation}%
where $\xi$, $\hat{\mu}_{2}$ and $\hat{\mu}_{3}$ are defined like
the previous section.

Varying the action (\ref{Act03}) with respect to $\xi (\rho)$
yields
\begin{equation}
\left( 1+2\hat{\mu}_{2}\xi +3\hat{\mu}_{3}\xi
^{2}\right) \frac{dN(\rho )}{d\rho }=0,
\label{eom1}
\end{equation}%
which shows that $N(\rho )$ should be a constant. We obtain the
below equation by Variation of the action (\ref{Act03}) with
respect to $h(\rho )$ and using $N(\rho )=1$ as
\begin{equation}
(n-1) h^{\prime} ({\beta}^{2}- h^{\prime 2}) +\rho h^{''}
{\beta}^{2} =0, \label{eom2}
\end{equation}%
Now, we can show that the vector potential can be written as
\begin{equation}
h(\rho)=-\sqrt{\frac{(n-1)}{2n-4}}\frac{q}{\rho^{n-2}} \Gamma(\eta
), \label{Amu2}
\end{equation}
where $q$ is  is related to the charge parameter and it is an integration constant and $\eta$ is
\begin{equation}
\eta =\frac{{(n-1)(n-2)q^{2}l^{2n-4}}}{2\beta ^{2}\rho^{2n-2}}.
\end{equation}
In Eq. (\ref{Amu2}), $\Gamma$ is the hypergeometric function that
we show its form here,
\begin{equation}
{_{2}F_{1}}\left( {%
\left[ \frac{1}{2},\frac{n-2}{2n-2}\right] ,\left[
\frac{3n-4}{2n-2}\right] ,-z }\right) =\Gamma(z).  \label{hyp}
\end{equation}
The hypergeometric function $\Gamma(\eta ){\rightarrow 1}$ as
$\eta \rightarrow 0$ ($\beta \rightarrow \infty $) and therefore
$h(\rho)$ of Eq. (\ref{Amu2}) reduces to the gauge potential of
Maxwell field. Variation with respect to $N(\rho )$ and
substituting $N(\rho )=1$ gives
\begin{equation}
\hat{\mu}_{3}\xi ^{3}+\hat{\mu}_{2}\xi
^{2}+\xi +\kappa =0,  \label{Eq4}
\end{equation}%
where
\begin{equation}
\kappa
=1-\frac{m}{\rho^{n}}+\frac{2(n-2)l^{2}\beta^{2}}{n(n-1)}[1-\sqrt{1+\eta}-\frac{\eta}{n-2}\Gamma(\eta
)]
\end{equation}%

\begin{figure}[htbp]
    \includegraphics[scale=0.8]{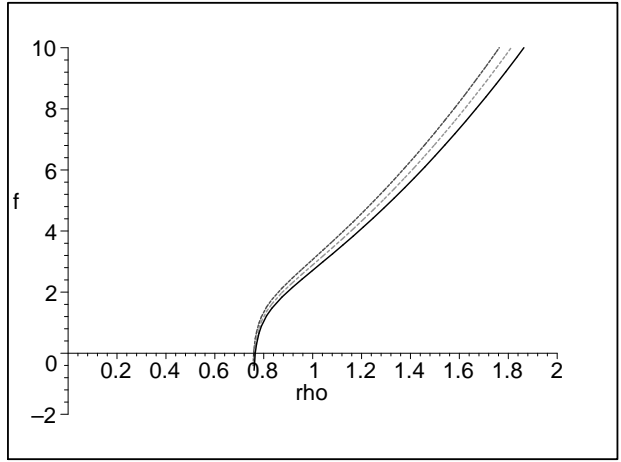}
    \caption{The overlay plot of $f(\rho)$ versus $\rho$ for $\hat{\mu}_{3}=0.008$
    (solid), $\hat{\mu}_{3}=0.0085$ (dotted) and $\hat{\mu}_{3}=0.009$ (dashed). Here, $l =1
    $, $\hat{\mu}_{2} = 0.1$, $q=10$, $\beta=50$ and $m=0.1$.}
    \label{BI3f01}
\end{figure}

\begin{figure}[htbp]
    \includegraphics[scale=0.8]{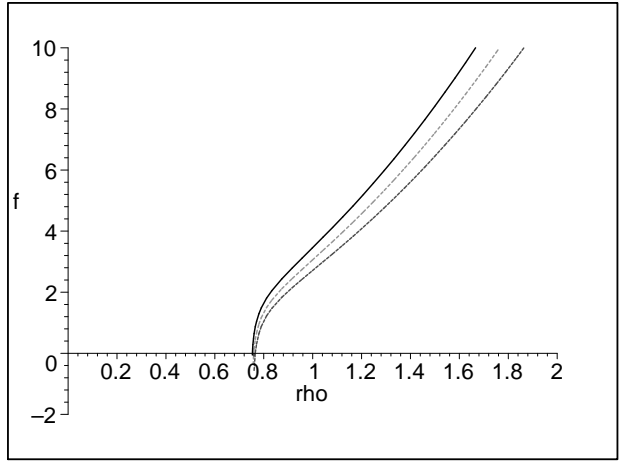}
    \caption{The overlay plot of $f(\rho)$ versus $\rho$ for $\hat{\mu}_{2} = 0.1$
    (solid), $\hat{\mu}_{2} = 0.102$ (dotted) and $\hat{\mu}_{2} = 0.104$ (dashed). Here, $l =1%
    $, $\hat{\mu}_{3}=0.009$, $q=10$, $\beta=50$ and $m=0.1$.}
    \label{BI3f02}
\end{figure}

After that, we can calculate the $f(\rho)$ function that leads the
same function as the $f(\rho)$ in the previous section (Eq.
\ref{Fr4}), but the value of $\kappa$ is different. We can see the
behavior of $f(\rho)$ function versus $\rho$ in Fig.
(\ref{BI3f01}) and (\ref{BI3f02}) that there is a curvature
singularity at $\rho=\rho_0$. We can find the same results as
obtained in the previous section for Quasi-Topological-Maxwell
gravity. Here, in the presence of a nonlinear electromagnetic
Born-Infeld field, we can see that the effect of $\beta$ is
negligible.

Again, here we look for curvature singularities. By using the
transforming (\ref{trans01}) and the metric (\ref{met02}) the
functions $\eta$, $h(r)$ and $\kappa $ becomes:
\begin{equation}
\eta =\frac{{(n-1)(n-2)q^{2}l^{2n-4}}}{2\beta
^{2}{(r^{2}+r_{+}^{2})}^{{2n-2}/2}}.
\end{equation}
\begin{equation}
h(r)=-\sqrt{\frac{(n-1)}{2n-4}}\frac{q}{{(r^{2}+r_{+}^{2})}^{{(n-2)}/2}}
\Gamma(\eta ), \label{Amu3}
\end{equation}
\begin{equation}
\kappa
=1-\frac{m}{{(r^{2}+r_{+}^{2})}^{n/2}}+\frac{2(n-2)l^{2}\beta^{2}}{n(n-1)}[1-\sqrt{1+\eta}-\frac{\eta}{n-2}\Gamma(\eta)]
\end{equation}
where $m$ is an integration constant which is related to the mass
of the space-time.

\begin{figure}[htbp]
    \includegraphics[scale=0.8]{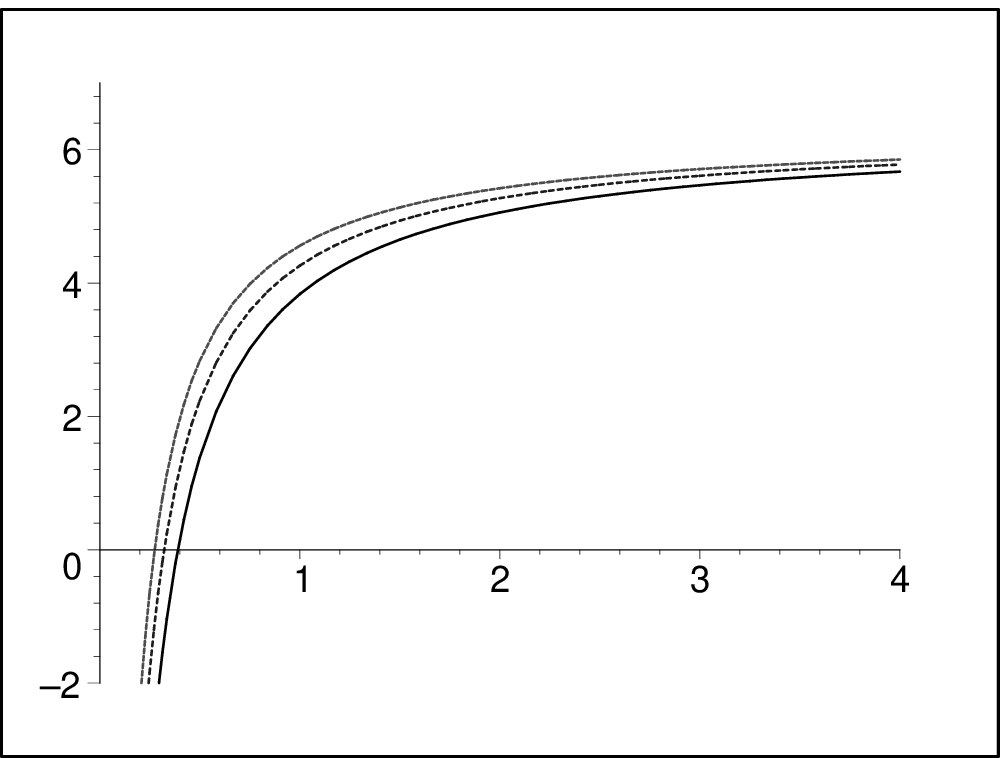}
    \caption{The overlay plot of $\delta$ versus $r_{+}$ for $\hat{\mu}_{3}=1$
    (solid), $\hat{\mu}_{3}=1.1$ (dotted) and $\hat{\mu}_{3}=1.2$ (dashed). Here, $l =1%
    $, $\hat{\mu}_{2} = 6$, $q=0.3$ and $\beta=0.01$.}
    \label{BI3dc01}
\end{figure}

\begin{figure}[htbp]
    \includegraphics[scale=0.8]{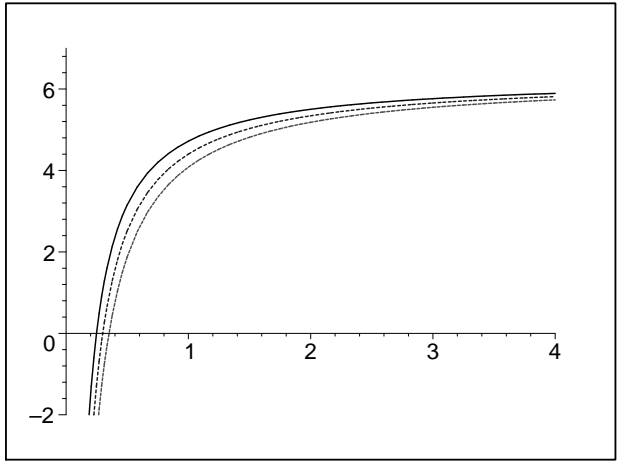}
    \caption{The overlay plot of $\delta$ versus $r_{+}$ for $\hat{\mu}_{2}=6$
    (solid) and $\hat{\mu}_{2}=9$ (dotted) and $\hat{\mu}_{2}=12$ (dashed). Here, $l =1%
    $, $\hat{\mu}_{3}= 1$, $q=0.3$ and $\beta=0.01$.}
    \label{BI3dc02}
\end{figure}

According the Eq. (\ref{limit}) up to Eq. (\ref{tau}) and
calculating the second derivative of $f(r)$, we plot $\delta$
versus the parameter $r_{+}$ that we show this in Fig.
(\ref{BI3dc01}) and (\ref{BI3dc02}). These plots show that by
increasing $r_{+}$, the value of the deficit angle increases.
Here, We can see the same conclusions as obtained in the previous
section, too. The deficit angle plots (Fig. \ref{BI3dc01},
\ref{BI3dc02}) show that $r_{+}$ increases as the $\hat{\mu}_{3}$
parameter increases, whereas for increasing the $\hat{\mu}_{2}$
parameter, $r_{+}$ decreases, but the $\beta$ effect is
negligible.

\section{Concluding REMARKS} \label{conclusion}
In this paper, we constructed magnetic solutions of the cubic
quasi-topological gravity in the presence of a linear Maxwell
field and a nonlinear Born-Infeld field. These solutions have no
horizon and calculations of geometric quantities showed the
solutions do not have curvature singularity. By using a suitable
radial transformation, we omitted change of signature and found a
conic singularity at $r=0$. Next, we investigated the effects of
different parameters on deficit angle and behavior of $f(r)$
function. In two sections, we considered the effects of
$\hat{\mu}_{2}$ and $\hat{\mu}_{3}$ parameters on metric function
and deficit angle in cubic quasi-topological action in the
presence of the linear and nonlinear electromagnetic field. We
found that the place of the root of the metric function was an
increasing function of the cubic quasi-topological parameter and a
decreasing function of the GB parameter. We obtained that the
$\beta$ parameter of Born-Infeld field do not have a significant
effect on the metric function and deficit angle. In the presence
of Maxwell field and Born-Infeld field, we saw that the metric
function and deficit angle have the same behavior. Therefore, we
found that the parameters that modified the behavior of the metric
function and the deficit angle graphs were cubic quasi-topological
and the GB parameters.

\begin{acknowledgements}
We would like to thank Payame Noor University and Jahrom
University.
\end{acknowledgements}

\end{document}